\documentclass[12pt]{article}
\usepackage{cite}
\usepackage[cmex10]{amsmath}
\usepackage{amssymb}
\usepackage{color}
\usepackage{algorithmic}
\usepackage{mdwmath}
\usepackage{mdwtab}
\usepackage{graphicx}

\def\R{\mathbb{R}}

\begin{document}
\title{Compartmental analysis of nuclear imaging data for the quantification of FDG liver metabolism}

\author{Valentina Vivaldi$^{1,4}$, Sara Garbarino$^{1,4}$, Giacomo Caviglia$^1$  \\ Michele Piana$^{1,4}$ and Gianmario Sambuceti$^{2,3}$\\
\small{$^1$Dipartimento di Matematica, Universit\`a di Genova, Italy} \\
\small{$^2$IRCCS San Martino IST, Genova, Italy}\\
\small{$^3$Dipartimento di Scienze della Salute, Universit\`a di Genova, Italy}\\
\small{$^4$CNR - SPIN, Genova, Italy}}

\maketitle

\begin{abstract}
This paper utilizes compartmental analysis and a statistical optimization technique in order to reduce a compartmental model describing the metabolism of labelled glucose in liver. Specifically, we first design a compartmental model for the gut providing as output the tracer concentration in the portal vein. This quantity is then used as one of the two input functions in a compartmental model for the liver. This model, in turn, provides as output the tracer coefficients quantitatively describing the effectiveness with which the labelled glucose is transported between the different compartments. For both models, the computation of the solutions for the inverse problems is performed by means of an Ant Colony Optimization algorithm. The validation of the whole process is realized by means of synthetic data simulated by solving the forward problem of the compartmental system.
\end{abstract}

\section{Introduction}
Positron Emission Tomography (PET) \cite{Ollinger} is an imaging technique capable of detecting picomolar quantities of a labelled tracer with temporal resolution of the order of seconds. FDG-PET is a PET modality in which [$^{18}$F]fluoro-2-deoxy-D-glucose (FDG) is used as a tracer to evaluate glucose metabolism and to detect diseases in many different organs \cite{Hays}. From a mathematical viewpoint, PET (and, specifically, FDG-PET) experiments involve two kinds of problems. First, signal processing techniques are applied to reconstruct the time dependence of location and concentration of tracer from the measured radioactivity. Second, these dynamic PET data can be processed to estimate physiological parameters that describe the functional behaviour of the inspected tissues. The present paper focuses on this second aspect and, specifically, examines FDG-PET data from liver (and gut), corrected for radioactive decay, to recover information on glucose metabolism. The synthetic data processed within the paper have been created mimicking the acquisition of a PET device for small animals (mice).

The analysis of tracer kinetics in the hepatic system, as in general in tissues whose activities have been measured with FDG-PET, is based on compartmental models \cite{Schmidt}. This conceptual framework identifies different compartments in the physiological system of interest, each one characterized by a specific (and homogeneous) functional role. As any other tracer, FDG is injected into the system with a concentration mathematically modeled by the so-called Time Activity Curve (TAC). The time dependent concentrations of tracer in each compartment constitute the state variables that can be determined from PET data. The time evolution of the state variables (the kinetics of the system) is modeled by a linear system of ordinary differential equations for the concentrations, expressing the conservation of tracer during flow between compartments. The (constant) micro-parameters describing the input/output rates of tracer for each compartment are called {\it exchange coefficients} or {\it rate constants}, represent the physiological parameters describing the system's metabolism, and are the unknowns to be estimated. 

Under the assumption that the exchange coefficients are given, the differential equations for the state variables in each compartment can be solved straightforwardly. If instead the exchange coefficients are unknown, we consider the formal expressions for the concentration time dependence involving the rate coefficients, and we compare them with the experimental data in order to optimize the unknown parameters. The data are given as total concentrations (e.g., inside liver) which are obtained by on drawing Regions Of Interest (ROIs) of the organ of interest at different time points and dividing the total activity by the corresponding volume. 

In addition to such general features, modelling of tracer flow inside liver implies a second difficulty, arising from physiology, in that the blood supply (and hence the tracer supply) comes from both hepatic artery and portal vein. While tracer concentration in the hepatic artery can be estimated by drawing ROIs on the left ventricle at different times, the portal vein is not accessible to ROIs. In the literature there have been many attempts to deal with this dual input problem \cite{Kudomi}. In this paper the input of tracer through the hepatic vein is estimated by the analysis of tracer flow in gut which is performed by means of an appropriate compartment modelling, where trapping of tracer is explicitly considered. The related experimental data are obtained from measurements of activity on ROIs drawn on the gut. Once the exchange coefficients have been determined, tracer concentrations in compartments inside gut are reconstructed, and the concentration of outgoing tracer through the hepatic vein is evaluated by application of a related balance equation.

From a methodological viewpoint, the inverse problem of determining the unknown exchange coefficients is solved by applying an Ant Colony Optimization (ACO) technique \cite{Dorigo,Socha}, which is a statistical optimization algorithm inspired by evolutionary strategies. In our application we minimize the discrepancy between the experimental concentrations and the already mentioned analytical forms provided by the solutions of the forward problems, taking also into account the blood fraction with which the vascular system of the liver is fed up. 

The paper is organized as follows. In Section 2 the systems of differential equations modelling the tracer flow inside gut and liver are described and formally solved for the concentrations in terms of the rate constants. Section 3 describes ACO. In Section 4 the exchange coefficients are determined starting from synthetic data. Our conclusions are offered in Section 5.

\section{The direct problem}

\begin{figure}[h!]
\hspace*{\fill}
\includegraphics[scale=0.4]{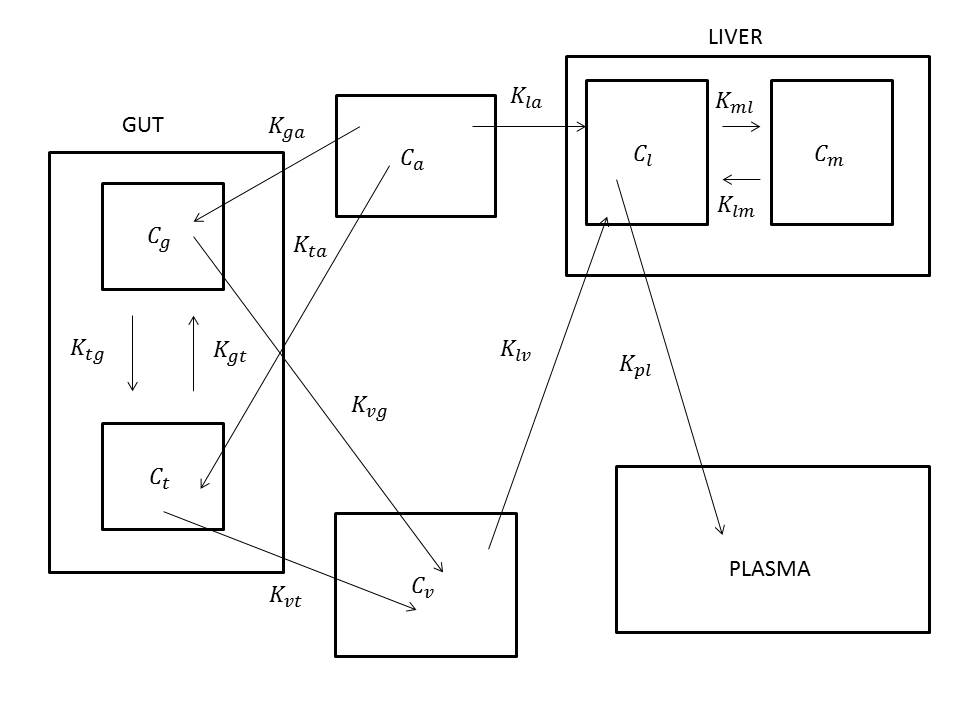}
\hspace*{\fill}%
\caption{The compartmental model of liver and gut. From left to right: the gut system, arterial and venus blood compartments, liver system, and an output blood compartment}\label{fig:gut-liver}
\end{figure}

The compartmental models used in this paper are described in Figure \ref{fig:gut-liver}. Tracer from arterial compartment enters the gut and the liver systems. It leaves the gut system, goes first into the portal vein and then into the liver. Capital $C$ denotes each concentration (Bq min$^{-1}$) and the related suffixes identify the corresponding compartment. In particular $a$ and $v$ refer to arterial and venous compartment. Gut includes the two compartments of concentration $C_g$ and $C_t$ of free and trapped tracer, while liver contains the two compartments of concentration $C_l$ and $C_m$ of free and metabolized tracer, respectively. The constant exchange coefficients between compartments in contact are denoted as $K_{ab}$ (min$^{-1}$), where the suffixes $a$ and $b$ denote the target and source compartment, respectively.  For example, $K_{ga}$ is the rate constant of tracer carried ``to'' the free gut tissue $g$ ``from'' the arterial blood $a$. We assume $K_{ab} \geq 0$ for all cases.
The usual conditions for compartmental analysis hold, e.g., tracer is uniformly distributed in each compartment at each instant, diffusive effects are neglected and physiological processes are in a steady state. Moreover, the kinetic process in the system is initialized by the TAC $C_{a}$ representing the tracer concentration in arterial blood. Accordingly, the initial values of the remaining concentrations vanish. In the following subsections we analyze tracer flow from gut, portal vein and the liver. The output from gut is the input for $C_v$ and $C_v$ is one of the two inputs of the liver.

\subsection{Gut}

The state variables of the two-compartment model adopted for the gut in this paper are the free tracer concentration in tissue ($C_{g}$) and the trapped tracer concentration ($C_{t}$). Conservation of tracer exchanged between compartments leads to the following system of linear ordinary differential equations with constant coefficients:
\begin{equation}\label{eq:system_gut}
\begin{cases}
\dot{{C}_{g}}=-({K}_{tg}+{K}_{vg}) {C}_{g} + {K}_{gt} {C}_{t} +{K}_{ga}  {C}_{a}\\
\dot{{C}_{t}}={K}_{tg}{C}_{g}-({K}_{gt}+{K}_{vt}){C}_{t}+{K}_{ta}{C}_{a},
\end{cases}
\end{equation}
with initial conditions $C_{g}(0)=C_{t}(0)=0$ and where the dependence on time is implicit.\\

Let us define
$$ a={K}_{tg}+{K}_{vg}, \, \, \, \, \, \, \, b={K}_{gt},\, \, \, \, \, \, \, \ c={K}_{tg},\, \, \, \, \, \, \, \ d={K}_{gt}+{K}_{vt}.$$
Then the solution of the Cauchy problem (\ref{eq:system_gut}) is given by
\begin{equation}\label{eq:gut_sol}
C_g=c_1 b E_1+c_2(d+\lambda_2)E_2, \, \, \, \, \, \, \, \, \, 
C_t=c_1(a+\lambda_1)E_1+c_2cE_2
\end{equation}
where
$$ E_i(t; \lambda_i)=\int_{0}^t e^{\lambda_i(t-\tau)}C_a(\tau)\, d\tau\, = e^{\lambda_it} * C_a\, \, \, \, \, \, \, \, \, i=1,2,$$
$*$ denotes the convolution operator, \begin{equation}\label{eq:rel_lambda}
{\lambda}_{1,2}= {-(a+d) \pm \sqrt{{(a+d)}^{2}-4(ad-bc)}  \over  2},
\end{equation} are the eigenvalues of the coefficients matrix, and the constants $c_1$ and $c_2$ are defined as
\begin{equation}\label{eq:eqc}
c_1={(d+\lambda_2)K_{ta} -cK_{ga} \over (a+\lambda_1)(d+\lambda_2)-bc}, \, \, \, \, \, \, \, \, \, c_2={(a+\lambda_1)K_{ga}-bK_{ta} \over (a+\lambda_1)(d+\lambda_2)-bc},
\end{equation}
under the assumption of non vanishing denominator. 

\subsection{Portal vein}
We represent the portal vein as a compartment with tracer concentration $C_v$ and volume $V_v$ placed between the gut and liver systems. The input flow from gut is described as $F_{vt} C_t + F_{vg} C_g $ where $F$ denotes the flow of carrier fluid per unit time, in analogy with the notation for the exchange coefficients. The outgoing flow of tracer towards liver is $(F_{vt} + F_{vg})C_v,$ where conservation of the carried fluid is explicitly considered. The balance equation for tracer takes the form 
$$V_v\dot{C_{v}}= F_{vt}C_t+F_{vg}C_g-(F_{vt}+F_{vg})C_v.$$ 
Division by $V_v$ yields
\begin{equation}\label{eq:cons_tr}
\dot{C_{v}}=K_{vt}C_t+ K_{vg}C_g-(K_{vt}+K_{vg})C_v,
\end{equation}
with $K_{vt} = {F_{vt} \over V_v}$ and $K_{vg} = {F_{vg} \over V_v}$.
Solving the differential equation (\ref{eq:cons_tr}) for vanishing initial value of $C_v$ leads to
\begin{equation}\label{eq:Cv}
C_v=\int_{0}^{t}e^{-(K_{vt}+K_{vg})(t-\tau)}(K_{vt}C_t(\tau)+K_{vg}C_g(\tau))\, d\tau.
\end{equation}
This provides tracer concentration of portal vein.\\

\subsection{Liver}
We refer to Figure \ref{fig:gut-liver}, with the observation that there is no flow of tracer from the metabolized compartment to the blood circulation out of the liver, that is $K_{bm} = 0$. Following the analogy with the analysis of the gut system, we find that the concentrations $C_l$ and $C_m$ inside the liver solve the system of linear ordinary differential equations
\begin{equation}\label{eq:eqfegatoKpm}
\begin{cases}
{\dot{C}}_{l}=-(K_{bl}+K_{ml})C_l+K_{lm}C_m+\sigma_{l}\\
{\dot{C}}_{m}=K_{ml}C_l-K_{lm}C_m,
\end{cases}
\end{equation}
where $\sigma_{l}$=$K_{la}C_a$+$K_{lv}C_v$ is known.\\

The solution of the Cauchy problem with the vanishing initial conditions takes the form
\begin{equation}\label{eq:liver_sol}
\begin{cases}
C_l= {K_{lm} \over \lambda_1-\lambda_2 }(E_1-E_2)+ {\lambda_1 E_1-\lambda_2 E_2 \over \lambda_1-\lambda_2}\\
C_m= {K_{ml} \over \lambda_1-\lambda_2 }(E_1-E_2),
\end{cases}
\end{equation}
with
$$\lambda_{1,2}= {-(K_{bl}+K_{ml}+K_{lm}) \pm \sqrt{{(K_{bl}+K_{ml}+K_{lm})}^2 - 4 K_{lm}K_{bl}} \over 2}.$$

\subsection{Optimization problems}
The model equations (\ref{eq:gut_sol}) and (\ref{eq:liver_sol}) describe the time behavior of the tracer concentration in the gut and the liver systems in terms of the TAC of the arterial concentration and exchange coefficients. Given such equations we have to determine the tracer coefficients by utilizing measuraments of the total tracer concentrations in gut and liver provided by nuclear imaging and applying an optimization scheme for the solution of the inverse problem.

Analysis of PET images on gut ROIs provides the total tracer concentration $R_1$ in gut at different time points. Similarly, ROIs on the liver provide the total tracer concentration $R_2$ in liver at the same time points. The optimization problems we have to solve are the ones to minimize
\begin{equation}\label{opt-1}
{\cal{C}}_1 = {\| R_1 - (1-V_a)(C_g + C_t) - V_a C_a \| }_2^2
\end{equation}
and
\begin{equation}\label{opt-2}
{\cal{C}}_2 = {\|R_2 - (1-V_b)(C_l + C_m) - V_b ({1 \over 4} C_a + {3 \over 4}C_v) \|}_2^2,
\end{equation}
where $V_a$ and $V_b$ are the blood volume fractions inside gut and liver respectively (in the following we will assume $V_a=0$ and $V_b = 0.4$). We have also taken the estimate that the ingoing flux of venous blood is about 3/4 of total flux while arterious blood is the remaining 1/4 \cite{Munk}.

\section{Ant Colony Optimization}
The minimization of the functionals ${\mathcal{C}}_1$ and ${\mathcal{C}}_2$ is realized by means of an Ant Colony Optimization (ACO) scheme \cite{Dorigo}. ACO is a statistical-based optimization method developed in the 1990s with the aim of providing a reliable although not optimal solution to some non-deterministic polynomial-time hard combinatorial optimization problems. While an ant is going back to the nest after having taken some food, it releases a pheromone trace that serves as a trail for next ants, which are able to reach food detecting pheromone. Since the pheromone decays in time, its density is higher if the path to food is shorter and more crowded; on the other hand, more pheromone attracts more ants and at the end all ants follow the same trail. This behavior is paraphrased in ACO identifying the cost functional with the length of the path to food, and the pheromone traces with a probability density which is updated at each iteration depending on the value of the cost function for a set of states. 

In practice, at each iteration, the cost function is evaluated on a set of $P$ admissible states, and the states are ordered according to increasing values of the cost function. Then ACO defines a probability distribution which is more dense in correspondence with the cheaper states and, on its basis, $Q$ new states are extracted. A comparison procedure identifies the new best $P$ states which form the next set of states.
Formally the starting point of the algorithm is a set of $P$ states
\begin{equation}\label{aco-3}
B:=\{\mathbf{U}_k=(u_{1,k},\dots,u_{N,k})\},\ \
\end{equation}
such that
\begin{equation}\label{aco-4}
\mathbf{U}_k\in S\subset  \R^N, \ \ \ k=1,\dots,P
\end{equation}
that are ordered in terms of growing cost, namely, $\mathcal{C}({\mathbf{U}}_1)\leq\dots\leq\mathcal{C}({\mathbf{U}}_P)$.
Next, for each $j=1,\dots,N$ and $i=1,\dots,P$, one computes the parameters
\begin{equation}\label{aco-5}
m_{i,j}=u_{j,i}, \ \ \ \ s_{i,j}=\frac{\xi}{P-1}\sum_{p=1}^P|u_{j,p}-u_{j,i}|, 
\end{equation}
and defines the probability density function
\begin{equation}\label{aco-6}
\mathcal{G}_j=\sum_{i=1}^Pw_i\mathcal{N}_{[m_{i,j},s_{i,j}]}(t),
\end{equation}
with $i=1,\dots,P$ and $\xi,q$ real positive parameters to be fixed and $w_i=\mathcal{N}_{[1,qP]}(i)$.
By sampling $S$ $Q$ times with $\mathcal{G}_j$, the procedure generates $Q$ new states $\mathbf{U}_{P+1},\dots,\mathbf{U}_{P+Q}$, enlarging the set $B$ to $\tilde{B}=\{\mathbf{U}_1,\dots,\mathbf{U}_{P+Q}\}.$ If $\mathbf{U}_{k_1},\dots,\mathbf{U}_{k_Q}$ are the $Q$ states of $\tilde{B}$ of greater cost, the updated $B$ is defined as 
\begin{equation}\label{aco-7}
B=\tilde{B}\setminus\{\mathbf{U}_{k_1},\dots,\mathbf{U}_{k_Q}\}.
\end{equation}
This procedure converges to an optimal solution of the problem by exploiting the fact that the presence of the weights $w_i$, in the definition of $\mathcal{G}_j$, gives emphasis to solutions of lower costs since $w_1>\dots>w_P.$ This fact, associated with the influence that a proper choice of parameters $\xi$ and $q$ has on the shape of the Gaussian functions, determines the way in which the method tunes the impact of the worse and best solutions.
The algorithm ends when the difference between any two states of $B$ is less than a pre-defined quantity or when the maximum allowable number of iterations is reached. The initial set $B$ of trial states is chosen by sampling a uniform probability distribution. 

The implementation of ACO for the optimization of the exchange coefficients in ${\cal{C}}_1$ and ${\cal{C}}_2$ is based on the following steps:
\begin{enumerate}
\item The four ACO parameters are fixed as follows. $P$ and $Q$ are chosen as in \cite{Giorgi}. Specifically, $P$ is a multiple of the number of coefficients to optimize plus one and
\begin{equation}\label{aco-8}
Q = \left[ \frac{P}{2} \right] + 1,
\end{equation}
where $[\cdot]$ indicates the floor; $q$ and $\xi$ are fixed searching for a trade off between the risk of a solution space of a too high complexity and the risk of a too high computational demand.
\item The values of the tracer coefficients are initialized to six random numbers picked up in the interval $(0,1)$.
\end{enumerate}
In the following section we show how this statistics-based compartmental analysis works in the case of synthetic data. In this specific application the advantages of ACO with respect to deterministic optimization are that it does not suffer local minima and singularities in the functional gradient.

\section{Numerical experiments}

\begin{figure}[h!]
\hspace*{\fill}
\includegraphics[scale=0.5]{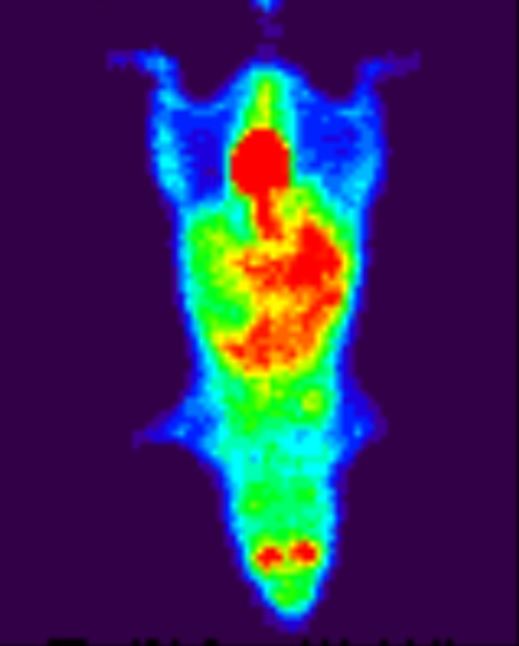}
\hspace*{\fill}%
\caption{A real micro-PET image of a mouse in a FDG experiment. This represents the typical nuclear medicine data for real application.}\label{fig:mouse}
\end{figure}

In order to validate the compartmental model described in Section 2 and the numerical method for its reduction described in Section 3 we have utilized synthetic data simulated by mimicking the behavior of a real micro-PET system (see picture in Figure \ref{fig:mouse}).
\begin{enumerate}
\item The input function $C_a$ has been constructed by using the Golish function \cite{Golish}
\begin{equation}\label{golish}
\theta(t) = C_{max} 
\left(\frac{et}{\alpha \beta}\right)^{\alpha} e^{-t/\beta} + C_0 e^{-t/\tau},
\end{equation}
where the parameters $(C_{max},\alpha,\beta,C_0,\tau)$ have been determined by fitting this model against experimental values obtained by means of a real experiment performed under very controlled conditions;
\item we have chosen a set of values for the tracer coefficients of gut system, as shown in Table \ref{table:results}, and we have solved the corresponding forward problem by means of the equations (\ref{eq:system_gut});
\item the resulting values for $C_g$ and $C_t$ have been sampled at the time points typical of real micro-PET scanners and perturbed by means of realistic Poisson noise
\end{enumerate}

This first part of the validation process allowed the simulation of the synthetic data to use as input of the ACO algorithm for the reduction of the gut compartmental model. Specifically, ACO has been initialized by six random values in $(0,1)$, and then the algorithm has been run 20 times in order to study the stability of the method. The results are in Table \ref{table:results} in which we show the average values and the corresponding standard deviations over the 20 realizations, for the six coefficients. 

Using the average values of $K_{vt}$ and $K_{vg}$ in equation (\ref{eq:Cv}) one can easily obtain the tracer concentration for the blood in output from the gut. 
This same $C_v$ is one of the input functions for the compartmental model of the liver. We could therefore simulate a set of synthetic data for this model by means of the same procedure as before; the forward problem is now the one to determine $C_l$ and $C_m$ by means of equation (\ref{eq:liver_sol}). The resulting concentrations have been sampled and affected by means of Poisson noise. Again ACO has been applied 20 times with random initialization. The ground truth of tracer coefficients, averages and standard deviations for the reconstructed values are shown in Table \ref{table:results}. 

\begin{table}
\begin{center}
\begin{tabular}{c|c|c|c|c|c|c}
\hline
GUT & $K_{ga}$ & $K_{ta}$ & $K_{gt}$ & $K_{tg}$ & $K_{vt}$ & $K_{vg}$ \\
\hline
g.t. & $0.10$ & $0.003$ & $0.02$ & $0.10$ & $0.003$ & $0.30$ \\
\hline
mean & $0.10$ & $0.003$ & $0.02$ & $0.11$ & $0.003$ & $0.33$ \\
\hline
std  & $0.03$ & $0.001$ & $0.01$ & $0.01$ & $0.002$ & $0.01$ \\
\hline\hline
\end{tabular}
\vspace{1cm}
\begin{tabular}{c|c|c|c|c|c}
\hline\hline
LIVER & $K_{la}$ & $K_{lv}$ & $K_{ml}$ & $K_{lm}$ & $K_{bl}$ \\
\hline
g.t. & $1.6$ & $0.02$ & $0.5$ & $0.2$ & $1.3$ \\
\hline
mean & $1.99$ & $0.02$ & $0.55$ & $0.21$ & $1.25$ \\
\hline
std  & $0.11$ & $0.01$ & $0.03$ & $0.01$ & $0.01$ \\
\hline
\end{tabular}
\end{center}
\caption{Validation of the ACO-based method for the gut-liver compartmental model: ground truth (g.t.) values for the tracer coefficients; mean and standard deviation (std) of the resulting simulation problems.}
\label{table:results}
\end{table}

\section{Conclusions}
This paper describes a compartmental model for the study of FDG metabolism in the hepatic system. The model is validated by using synthetic data simulated by mimicking the acquisition process of a real micro-PET system for mice. The main difficulty in the compartmental analysis of nuclear data for the liver is in the fact that this organ presents two input blood functions and one of the two (the portal vein) is difficult to be determined. The reason for this is that the portal vein is invisible to PET analysis. Our approach solves this issue by setting up a separate compartmental analysis of data for the gut, where the tracer concentration in the portal vein is determined as solution of an inverse problem. Such solution is then utilized as input function for a second compartmental model, this time describing the tracer kinetics in liver. As a final outcome of the approach we are able to quantitatively determine the tracer coefficients modeling the FDG metabolism in the hepatic system. From a computational viewpoint, we prove that statistical optimization based on an Ant Colony Optimization algorithm is very effective for model reduction. The next step of this analysis will be its validation in the case of real measurements acquired by means of a micro-PET system. The application of this approach may concern the assessment of the effectiveness of specific chemical compounds able to inhibit the re-metabolization of FDG. These pharmacological tools are important in oncological applications since they could contrast the access to glucose for malignant cells. 


\end{document}